\documentclass[aps,pra,twocolumn,showpacs,superscriptaddress]{revtex4}

\usepackage{hyperref}
\usepackage{amsmath}
\usepackage{amssymb,dsfont}
\usepackage[small]{subfigure}
\usepackage[dvips]{graphicx,color}

\newcommand{\imu}{\textrm{i}}
\newcommand{\myfigwidth}{0.85\columnwidth}

\begin{document}
\newcommand{\mysize}{\scriptscriptstyle}

 \title{Molecular heat pump for rotational states}

 \author{C.~Lazarou}
 \affiliation{Department of Physics and Astronomy,
   University of Sussex, Falmer, Brighton, BN1 9QH,
   United Kingdom}
 \affiliation{Department of Physics, Sofia University, James Bourchier
   5 Boulevard, 1164 Sofia, Bulgaria}

 \author{M.~Keller}
 \affiliation{Department of Physics and Astronomy,
   University of Sussex, Falmer, Brighton, BN1 9QH,
   United Kingdom}

 \author{B.M.~Garraway}
 \affiliation{Department of Physics and Astronomy,
   University of Sussex, Falmer, Brighton, BN1 9QH,
   United Kingdom}

\date{\today.}

 \begin{abstract}
   In this work we investigate the theory for three different
   uni-directional population transfer schemes in trapped
   multilevel systems which can be utilized to cool molecular
   ions.  The approach we use exploits the laser-induced coupling
   between the internal and motional degrees of freedom so that
   the internal state of a molecule can be mapped onto the motion
   of that molecule in an external trapping potential.  By
   sympathetically cooling the translational motion back into its
   ground state the mapping process can be employed as part of a
   cooling scheme for molecular rotational levels.  This step is
   achieved through a common mode involving a laser-cooled atom
   trapped alongside the molecule. For the coherent mapping we
   will focus on adiabatic passage techniques which may be
   expected to provide robust and efficient population transfers.
   By applying far-detuned chirped adiabatic rapid passage pulses
   we are able to achieve an efficiency of better than 98\% for
   realistic parameters and including spontaneous emission.  Even
   though our main focus is on cooling molecular states, the
   analysis of the different adiabatic methods has general
   features which can be applied to atomic systems.
 \end{abstract}
 \maketitle

 \section{Introduction} \label{sec:intro}

 The emerging field of cold molecules is a very vibrant
 topic in physics and physical chemistry. The considerable
 interest in this topic is related to the properties of cold
 molecules and their many potential applications.  Cold
 molecules have been identified as attractive systems for
 ultrahigh-resolution spectroscopy
 \cite{highresSpec,Schuster09}, quantum information
 processing \cite{qip}, for developing new time standards
 and testing fundamental physical theories such as the time
 variation of physical constants
 \cite{schillingfundconst,ChengChin09}, the existence of a
 dipole moment of the electron \cite{eldipolemoment}, and
 for the measurement of parity violation \cite{parity}. In
 chemistry, cold molecules are essential tools to explore
 quantum-mechanical effects in chemical reactions.  In
 contrast to atoms, molecules have a very complicated level
 structure that consists of vibrational and rotational
 states as well as electronic levels. This abundance of
 states is the main obstacle for direct laser cooling of
 molecules. Usually, molecules do not provide the closed
 transitions required for cooling and non-destructive
 state-selective detection. This makes it impossible to
 perform direct spectroscopic measurements on
 \textit{single} molecules---a standard technique in atomic
 physics. Additional complications result from the small
 energy differences between the rotational levels, leading
 to a thermal distribution of the population over the
 molecules' ro-vibrational states.

 Despite important achievements \cite{buffergas, starkdec, optstarkdec},
 the control of molecular states never caught up with that of atomic
 systems.  However, there has been remarkable progress in the synthesis of
 ultra-cold alkali dimers from samples of ultra-cold atoms; see e.g.,
 Refs.\ \cite{photoassociation,Pillet08,feshbach}.  Furthermore, methods
 which enable the preparation of more diverse (e.g.\ polyatomic) cold
 molecular species in their vibrational ground-states have been
 successfully demonstrated.  These methods include: supersonic beam
 expansion followed by Stark deceleration \cite{starkdec}, optical Stark
 deceleration \cite{optstarkdec}, electrostatic velocity selection
 \cite{velocityselection}, collisional cooling in crossed molecular beams
 \cite{crossedbeam} and buffer gas cooling \cite{buffergas}.  The
 vibrationally cold (but rotationally hot) states that result will be taken
 to be the starting point for the schemes described in this paper.  The
 experimental advances which have enabled the production of these cold
 molecular states have inspired theoretical investigations of the cooling
 of molecules by laser pulses \cite{mollasercooling1, mollasercooling2,
   Kosloff1, Kosloff2, Kosloff3} or even by coupling molecules to an
 optical cavity \cite{cavitycooling}.  Bartana et al.\ \cite{Kosloff1} used
 the electronic excited-state as a heat reservoir in order to cool the
 vibrational states of the electronic ground-state by means of short,
 shaped, laser pulses. In later work \cite{Kosloff2} they employed state
 selective optical pumping, hiding the target state in a dynamically
 trapped state. Through this, Bartana et al.\ achieved a vibrational ground
 state population of 97\% after only 25 vibration periods. A related
 scheme, investigated by Schirmer \cite{mollasercooling1}, increased the
 vibrational ground-state population to a similar level.  These efforts to
 cool the internal degrees of freedom focus on the widely spaced
 vibrational states of molecules.  In Ref.\ \cite{Kosloff3} Bartana et al.\
 investigated the possibility of cooling the \emph{rotational} degrees of
 freedom in a simplified model employing the same techniques. However, even
 though the results of their calculations are very promising, the model has
 limited application as it neglects the vibrational degree of freedom of
 the molecule.
  
 In contrast to neutral molecules, ionized molecules can be sympathetically
 cooled by trapping them alongside atomic ions in a Paul trap. Under these
 conditions, state sensitive ultra-cold chemical reactions have been
 measured \cite{drewsenchemie1,Staanum08} and high resolution spectroscopy
 has been demonstrated on small ensembles \cite{schillerspectro}. Despite
 these achievements, the rotational degrees of freedom could not be
 controlled, but they led to new laser cooling schemes for the internal
 states of molecules \cite{vogelius1, vogelius2} exploiting the unique
 properties of these systems.

 In the present paper, we show that by means of purely coherent
 manipulations of internal states (i.e.\ rotational states), and by using
 sympathetic cooling of motional states, single molecular ions can be
 cooled close to their motional and rotational ground-state.  The internal
 vibration of the molecule needs to be initially cold, in order for the
 final state to be cold in \textit{all} its degrees of freedom.  This
 initial state can be achieved using the existing methods described above.
 Then the cooling of the internal molecular state is achieved in three
 steps: first a laser cooled atomic ion is trapped alongside the molecule
 and a common mode of vibration is used to prepare the molecule in its
 motional ground-state. Next, using adiabatic passage methods
 \cite{Melinger94,Chelkowski95,Chelkowski97,Malinovsky2001,%
   Grischkowsky1975,Yatsenko99,Rickes2000,Vitanov2001,Rice,Coulston1992},
 the thermal internal state of the molecule can be mapped onto the
 molecule's motion in an external trapping field with high fidelity.
 During this mapping process the internal (rotational) states are
 transferred to the ground-state whereas the molecule's motion is excited.
 Finally, the molecule's motion is sympathetically cooled back into its
 ground-state using the common mode with a cooled atomic ion. By doing this
 without exciting other degrees of freedom, a molecule which is
 vibrationally, rotationally, and translationally cold can be obtained. The
 overall process is a kind of molecular ``heat pump'': the heat energy in the
 rotational degree of freedom is transferred by the coherent processes to
 the motional degree of freedom.  This heat energy is in turn transferred
 to the environment by means of conventional cooling techniques, e.g.
 side-band cooling: this ensures that the process is uni-directional.

 Vogelius et al.\ \cite{vogelius1} investigated the cooling of molecular
 ions by coupling a single rotational state to the motion of the ion. By
 means of black-body radiation the population is pumped into the rotational
 ground-state resulting in a ground-state population of about 80\% after a
 cooling time of the order of minutes. In this paper we focus on the
 cooling process by employing techniques from coherent control providing
 much faster and more efficient cooling. The result is a robust and highly
 efficient cooling process which enables the deterministic manipulation of
 the internal states of molecules. We will examine adiabatic passage
 schemes which exhibit high fidelity in conjunction with relaxed
 requirements on the experimental parameters compared to direct Raman
 transitions. With cooling times of the order of milliseconds, and final
 ground-state populations of more than 92\%, the proposed scheme provides a
 fast and efficient method for preparing molecules in their ro-vibrational
 ground-state.

 Even though we focus here on the cooling of molecular rotational levels,
 the technique is also applicable to other multi-level systems. So the
 technique can also be employed in atoms with complicated level schemes, or
 more general ro-vibrational states of molecules.

 The paper is organized as follows: in Section \ref{sec:2}, we present the
 model used for our calculations. To find the best adiabatic passage
 process for our application we compare the results of numerical
 simulations for stimulated Raman adiabatic passage (STIRAP), Stark chirped
 Raman adiabatic passage (SCRAP) and chirped adiabatic rapid passage (CARP)
 in a $\Lambda$-type level system in Section \ref{sec:3}. We also choose
 parameter ranges which are relevant for a possible experimental
 implementation. Section \ref{sec:4} contains the results for an extended
 level scheme, and in Section \ref{sec:conclusion} we present our
 conclusions.

\section{The model} \label{sec:2}

The model we employ in this paper is based on the states of a quantum
mechanical rigid rotator which is a good approximation for the rotational
states of small diatomic molecules. However, the techniques described here
are applicable to most other level structures with the sole requirement
that allowed Raman transitions between the states involved exist.  In
general, the method can be applied to ro-vibrational states of molecules as
well as Zeeman and hyperfine levels.  In order to simplify the discussion,
and to avoid specializing to molecules with specific symmetries, we will
not apply specific selection rules to the Raman transitions involved.
Nevertheless, the results presented in this paper can be applied to a
particular system by imposing the specific selection rules for that case
with an appropriate relabelling of states.  For example, we could utilize
$\Delta J = 0, \pm 2$ for linear molecules.

\begin{figure}%[th]
  \begin{center}
    \includegraphics[width=0.85\columnwidth]{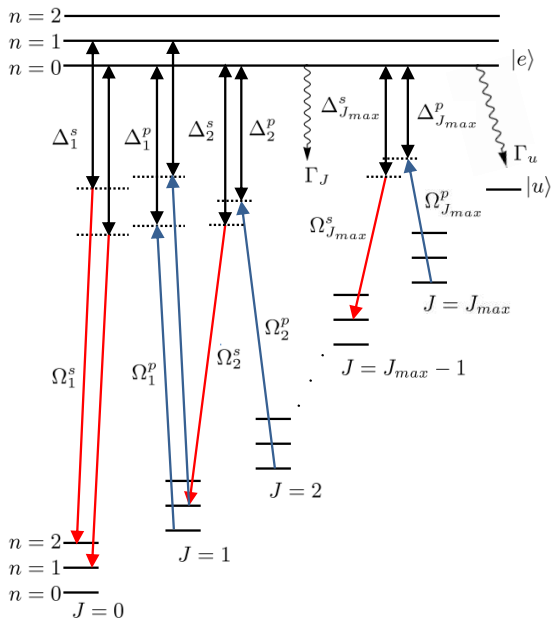}
    \caption[The molecular ion configuration.]{(Color online) 
      Overview of the
      molecular levels and the couplings which are included in the
      model.  Near-resonant laser pulses are driving the transitions
      between the excited levels $\vert e,n\rangle$ and the rotational
      states $\vert J,n\rangle$. The excited state decays towards the
      states $\vert J\rangle$ at a rate $\Gamma_{\mysize J}$ and
      towards the uncoupled state $\vert u\rangle$ at a rate
      $\Gamma_{\mysize u}$. The laser pulses have time-dependent Rabi
      frequencies $\Omega^{s,p}_J$ and are detuned from the relevant
      transition frequency by $\Delta^{s,p}_J$.}
    \label{fig:Levelsystem}
  \end{center}
\end{figure}

We take the energy $E_J$ of the rotational levels of the electronic ground
state to be $E_J=B\cdot J(J+1)$, with the rotational quantum number $J$ and
a rotational constant $B$. In order to limit the number of levels in our
calculations, only rotational levels up to a cut-off are considered, i.e.\
$J \leq J_{max}$.  For typical, light, diatomic molecules at room
temperature only rotational states with $J<20$ are significantly populated.
This decreases to below 10 states for rotational temperatures lower than
about 50K.  This kind of temperature can be easily achieved by supersonic
beam expansion \cite{scoles}.  In our scheme the levels $J$ are coupled by
laser pulses to an electronically excited state $\vert e\rangle$ (see Fig.\
\ref{fig:Levelsystem}). 
From the excited state the molecule can
spontaneously decay back into the electronic ground-state with a rate
$\Gamma_J$. To represent the decay of the excited state into levels outside
the system, e.g.\ rotational states with $J > J_{max}$ or vibrationally
excited states, an additional level $\vert u\rangle$ is also included in
the model. The excited state can decay into this uncoupled state with a
rate $\Gamma_u$.

The single molecule is trapped in a harmonic potential as provided by
rf-traps for molecular ions.  The quantized motion of the molecule in the
trapping potential gives rise to an equally spaced ladder of motional
states in addition to the internal states of the molecule. To take the
molecule's motion into account, we use the standard notation $\vert
i,n\rangle$ with $i$ representing the states $\vert J\rangle$, $\vert
e\rangle$ and $\vert u\rangle$. The quantum number describing the motional
state is $n$.  As discussed in the Introduction, in order to prepare the
molecule's motion in its ground-state, it is trapped alongside one or more
atomic ions which can be directly laser cooled.  The two types of ion form
a crystal-like structure due to their mutual Coulomb repulsion, which
enables sympathetic cooling of the molecule. In experiments with two types
of atomic ion, ground-state populations of better than 95\% have been
achieved \cite{sympcooling}.  In the molecular case, the minimal system
would be one trapped molecular ion and one trapped atomic ion. In this
case, we find that due to the frequency splitting of the motional modes,
one mode (COM or stretch mode) can be singled out and used for the proposed
scheme.  The other mode only imposes an additional limit on the laser pulse
length as we will discuss later in this paper.

The laser-molecule interaction, in the rotating wave approximation
\cite{Shore2}, is described by the following Hamiltonian \cite{note1}:
  \begin{equation} \label{eq:1}
  \begin{split}
    \hat{H}(t)=&\hbar\nu\hat{a}^\dagger\hat{a}
    +\sum_{J=1}^{J_{max}}\hbar\Delta_{\mysize J}^p(t)\vert e\rangle\langle e\vert
    \\&+\sum_{J=1}^{J_{max}}\hbar
    \left[\Delta_{\mysize J}^p(t)-\Delta_{\mysize J}^s(t)+\nu\right]\left\vert J-1\right\rangle\left\langle
      J-1\right\vert
    \\ &+\sum_{k=s,p}\left[\hat{H}^k_{c}(t)+\hat{H}^k_{r}(t)
      +\hat{H}^k_{b}(t)\right],
  \end{split}
\end{equation}
where $\hat{H}^{k}_{c}(t)$ represents the carrier resonance
\cite{Leibfried2003} for either the pump $(k=p)$, or the
Stokes-pulse $(k=s)$, with
\begin{equation} \label{eq:2}
  \begin{split}
    &\hat{H}^{p}_{c}(t)=\sum_{J=1}^{J_{max}}\frac{\hbar\Omega_{\mysize J}^{p}(t)}{2}\Big(\hat{\sigma}^{\mysize
      J}_{+}+\hat{\sigma}^{\mysize J}_{-}\Big), \\
    &\hat{H}^{s}_{c}(t)=\sum_{J=1}^{J_{max}}\frac{\hbar\Omega_{\mysize J}^{s}(t)}{2}\Big(\hat{\sigma}^{\mysize
      J-1}_{+}+\hat{\sigma}^{\mysize J-1}_{-}\Big).
  \end{split}
\end{equation}
The corresponding red sideband transitions in Eq.\ \eqref{eq:1} are the $\hat{H}^k_{r}(t)$,
which are given by
\begin{equation} \label{eq:3}
  \begin{split}
    &\hat{H}^{p}_{r}(t)=\sum_{J=1}^{J_{max}}\frac{\hbar\eta^p_{\mysize J}\Omega_{\mysize
      J}^{p}(t)}{2}\Big(\hat{\sigma}^{\mysize J}_{+}
  \hat{a}+\hat{\sigma}^{\mysize J}_{-}\hat{a}^\dagger\Big), \\
  &\hat{H}^{s}_{r}(t)=\sum_{J=1}^{J_{max}}\frac{\hbar\eta^s_{\mysize J}\Omega_{\mysize
      J}^{s}(t)}{2}\Big(\hat{\sigma}^{\mysize J-1}_{+}
  \hat{a}+\hat{\sigma}^{\mysize J-1}_{-}\hat{a}^\dagger\Big),
  \end{split}
\end{equation}
and for the first blue sideband transitions the
$\hat{H}^{k}_{b}(t)$ are given by
\begin{equation} \label{eq:4}
  \begin{split}
    &\hat{H}^{p}_{b}(t)=\sum_{J=1}^{J_{max}}\frac{\hbar\eta^p_{\mysize J}\Omega_{\mysize
        J}^{p}(t)}{2}\Big(\hat{\sigma}^{\mysize J}_{+}
    \hat{a}^\dagger+\hat{\sigma}^{\mysize J}_{-}\hat{a}\Big),\\
    &\hat{H}^{s}_{b}(t)=\sum_{J=1}^{J_{max}}\frac{\hbar\eta^s_{\mysize J}\Omega_{\mysize
        J}^{s}(t)}{2}\Big(\hat{\sigma}^{\mysize J-1}_{+}
    \hat{a}^\dagger+\hat{\sigma}^{\mysize J-1}_{-}\hat{a}\Big).
  \end{split}
\end{equation} 
In Eqs. (\ref{eq:1}-\ref{eq:4}), the secular frequency of the molecule in
the external trapping potential is $\nu$ and $\eta^k_{\mysize J}$ are the
corresponding Lamb-Dicke parameters \cite{Leibfried2003}. The detuning
between the $k$-th laser frequency $\omega^k_{\mysize J}$ and the
transition frequency $\omega_{\mysize eJ}$ of the $\vert J\rangle
\rightarrow \vert e\rangle$ transition is $\Delta^p_{\mysize J}(t)=
\omega_{\mysize eJ}-\omega^p_{\mysize J}(t)$ and $\Delta^s_{\mysize
  J}(t)=\omega_{\mysize eJ-1} - \omega^s_{\mysize J}(t) -\nu$.  The raising
operator for the internal states is $ \hat{\sigma}^{\mysize J}_+=\vert
e\rangle\langle J\vert$ and the lowering operator is $\hat{\sigma}^{\mysize
  J}_-=\vert J\rangle\langle e\vert$.  The creation and annihilation
operators for the motional number states $\vert n\rangle$ are
$\hat{a}^\dagger$, and $\hat{a}$ respectively.  Note that we work in an
interaction representation with explicit time-dependence removed, and keep
both the resonant and non-resonant couplings. The reason for this is that
in pursuing the adiabatic limit in Section \ref{sec:3} we will consider
``strong'' Rabi frequencies ($\Omega_J^k \sim \nu$) which do not allow us to
make a second RWA on the Hamiltonians for the sideband transitions.

We start our calculations with a molecule in an internal thermal state
such that it is already cooled in its internal vibrational mode (e.g.\ by the methods
mentioned in the Introduction) and such that the excited state $\vert e\rangle$ is
not populated. We also assume that the molecule's vibrational motion in the
trap has been cooled (e.g.\ by sympathetic sideband cooling
\cite{sympcooling}) so that only the manifold of rotational states $\vert J,0\rangle$
are populated.
The density matrix of this initial state is given by:
\begin{equation} \label{eq:5}
  \rho_{init} =\frac{1}{Z}\sum_{J=0}^{J_{max}}\left(2J+1\right)e^{-\beta B J(J+1)}\vert J,0\rangle\langle
  0,J\vert ,
\end{equation}
where $J_{max}$ is the cut-off introduced for the numerical calculations.
The normalization factor $Z$ is given by
$Z=\sum_{J=0}^{J_{max}}\left(2J+1\right)e^{-\beta B J(J+1)}$ with
$\beta=1/(k_{\mysize B}T)$, and with $T$ as the internal rotational temperature
of the molecule.

Starting from such an initial distribution, we will apply coherent control
techniques to transfer population between the different states. 
We aim to have a state mapping of the form
\begin{equation} \label{eq:6}
  \sum_{J=0}^{J_{max}}
  P_{J,0}\vert J,0\rangle\langle 0,J\vert\rightarrow\sum_{n=0}^{J_{max}}P_{0,n}\vert
  0,n\rangle\langle n,0\vert,
\end{equation}
where $P_{J,0} = P_{0,n} =e^{-\beta BJ(J+1)}/Z$ are the
populations of the initial states $\vert J,n=0\rangle$ and
the final target states $\vert J=0,n\rangle$ respectively. A
sequence of pulses will be used to map population in each
$J$-state to a corresponding $n$-state with $J=0$.

Throughout this work, we assume that the system is initially
prepared in a state given by Eq.\ (\ref{eq:5}), and derive
the requirements for achieving the state mapping in Eq.\
\eqref{eq:6}, i.e.\ after completion of a number of the
coherent pulse sequences. Thus we obtain a superposition of
just the motional states which can then be cooled to the
motional ground-state $\vert J=0, n=0\rangle$ by applying
the sympathetic cooling \cite{sympcooling}.  [The state
mapping \eqref{eq:6} can also be employed for
non-destructive state detection: e.g.\ by measuring the
initial thermal distribution (\ref{eq:5}).  By coupling the
electronic state of an atom trapped alongside the molecules
to its motion, the mapped state can be read out.]  For the
coherent mapping the key idea is to use pairs of pulses,
$\Omega^p_{\mysize J}(t)$ and $\Omega^s_{\mysize J}(t)$, to
induce population transfer between the states $\vert
J,n\rangle$ and $\vert J-1, n+1\rangle$, see Fig.\
\ref{fig:Levelsystem}.  For this step it is important to
have a resonance, so that additional states do not get
strongly involved and disturb the mapping. Here, the
resonance is arranged so that the quantity $J+n$ is
conserved at each step. (Although this is the simplest way
to make the mapping, it would not be the only way, as we
only require the transfer of population between unique pairs
of states.)

Repeating the $ J,n \longrightarrow J-1, n+1$ step $J_{max}$
times, where in each step $J$ is decreased by one, the
distribution of population can be moved to the $J=0$
motional states as in Eq.\ (\ref{eq:6}).  Since states with
$n\ne 0$ are involved in the intermediate steps, it is clear
that if we do not start in the motional ground-state the
final state need not be entirely $J=0$. However, when we
start in the motional ground-state (i.e.\ $n=0$), the
population transfer is uni-directional. Hence, the
population is transferred solely to the lower lying
rotational states.

In our analysis we numerically integrate the master equation for the
density matrix $\rho(t)$:
\begin{equation} \label{eq:7}
  \frac{d\rho(t)}{dt}= - \imu[\hat{H}(t),\rho(t)]+\hat{\mathcal{L}}(\rho(t))+\hat{\mathcal{L}}_{\mysize u}(\rho(t)).
\end{equation}
The Hamiltonian $\hat{H}(t)$ is given in Eq.\ \eqref{eq:1}.
The two Liouville terms $\hat{\mathcal{L}}(\rho(t))$ and
$\hat{\mathcal{L}}_{\mysize u}(\rho(t))$ describe the irreversible decay of the excited
electronic state $\vert e \rangle$
to the rotational levels $\left\vert J\right\rangle$ of the electronic and vibrational ground-state
\begin{equation} \label{eq:8}
  \hat{\mathcal{L}}(\rho(t))=-\sum_{J=0}^{J_{max}}\frac{\Gamma_{\mysize
      J}}{2}\Big(\rho(t)\hat{\sigma}^{\mysize J}_+\hat{\sigma}^{\mysize
    J}_-+\hat{\sigma}^{\mysize J}_+\hat{\sigma}^{\mysize
    J}_-\rho(t)-2\hat{\sigma}^{\mysize J}_-\rho(t)\hat{\sigma}^{\mysize J}_+\Big),
\end{equation}
and the decay out of the system, i.e.\ to the uncoupled state $\vert u\rangle$,
\begin{equation} \label{eq:9}
 \hat{\mathcal{L}}_{\mysize u}(\rho(t))=-\frac{\Gamma_{\mysize
     u}}{2}\Big(\rho(t)\hat{\sigma}^{\mysize u}_+\hat{\sigma}^{\mysize u}_-+\hat{
    \sigma}^{\mysize u}_+\hat{\sigma}^{\mysize u}_-\rho(t)-2\hat{\sigma}^{\mysize
  u}_-\rho(t)\hat{\sigma}^{\mysize
  u}_+\Big).
\end{equation}
In Eq.\ \eqref{eq:9}
the raising and lowering operators $\hat{\sigma}^{\mysize u}_{\pm}$ have the same form
as  $\hat{\sigma}^{\mysize J}_{\pm}$, with the substitution $\left\vert J\right\rangle\rightarrow\vert u\rangle$. 

We will assume that the laser pulses are  Gaussian, with a fixed
width $T$ and Rabi frequencies 
\begin{equation} \label{eq:10}
  \begin{split}
    &\Omega^p_{\mysize J}(t)=\Omega^p_{\mysize 0}e^{-(t-3\tau-(J_{max}-J)\tilde{\tau})^2/T^2},\\ \\
    &\Omega^s_{\mysize J}(t)=\Omega^s_{\mysize 0}e^{-(t-\tau-(J_{max}-J)\tilde{\tau})^2/T^2},
  \end{split}
\end{equation}
where $J=1,\cdots,J_{max}$.  The delay between the two pulses
$\Omega^s_{J}(t)$ and $\Omega^p_{J}(t)$ that drive the transition $\vert
J\rangle\rightarrow\vert J-1\rangle$ is $2\tau$, whereas the delay between
the $J$ and $J+1$ pulse pair is $\tilde{\tau}$. The corresponding detunings will
be either constants or have the form of time-dependent frequency chirps.  We
take all the Lamb-Dicke parameters to be the same, i.e.\ $\eta^{k}_{\mysize
  J}=\eta$, and for simplicity we assume that the decay rates are the same, i.e.\
$\Gamma_{\mysize J}=\Gamma_{\mysize u}=\Gamma$.  In order to compare the
three methods (STIRAP, SCRAP and CARP) in Section \ref{sec:3}
we characterize the population transfer efficiency
by a parameter $\epsilon$ representing
the total population of the rotational ground-state after the
transfer:
\begin{equation} \label{def:epsilon}
  \epsilon = \sum_{n=0}^{J_{max}}P_{0,n}(t=\infty),
\end{equation}
where $P_{0,n}(t=\infty)$ is the population of the $\vert J=0,n\rangle$
state after the transfer.  For an efficient state mapping, the ground-state
population ($J=0$) will have increased. However, if the initial state also
has some population in the $\vert J=0,n=0\rangle$ state, the efficiency
$\epsilon$ may also decrease due to laser-induced transfer out of $\vert
J=0,n=0\rangle$. Thus the definition \eqref{def:epsilon} is not only a
measure for the transfer efficiency, but also for the uni-directionality of the
mapping process. For ideal state mapping the efficiency measure reaches the
limit $\epsilon = 1$. In order to test the various passage methods in the
next Section, the initial state is taken to be a mixture of 70\%
rotationally excited states ($J > 0$) and 30\% ground-state ($J=0$)
population for each adiabatic method.  This approach will test how
uni-directional the scheme is.

\section{Adiabatic passage methods} \label{sec:3} 

\begin{figure}
  \begin{center}
    \includegraphics[width=\myfigwidth]{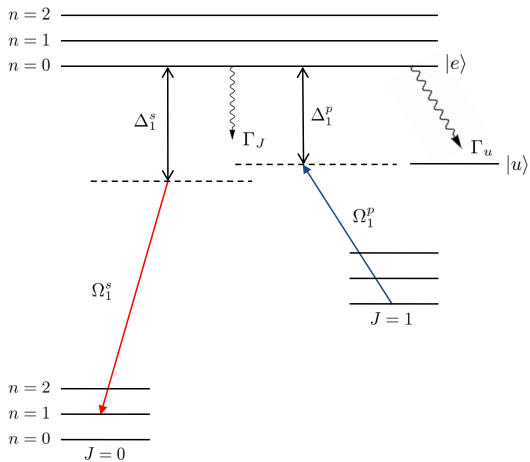}
    \caption[The $\Lambda$-configuration.]{(Color online) 
      The basic $\Lambda$-configuration
      used to investigate the transfer efficiency of various adiabatic
      passage schemes in Section \ref{sec:3}. The pump and Stokes pulses,
      $\Omega^p_{\mysize1}(t)$ and $\Omega^s_{\mysize0}(t)$ respectively,
      will induce a two-photon Raman transition (which may be chirped) from
      state $\vert n=0,J=1\rangle$ to state $\vert n=1,J=0\rangle$.  The
      best results are obtained if the initial population in $\vert
      n=0,J=0\rangle$ remains there.  The off-resonant intermediate states
      $\vert e,0 \rangle$ decays towards the two rotational states and
      towards the uncoupled state $\vert u\rangle$ at a rate
      $\Gamma_u$. 
      The decay rate $\Gamma_J$ represents decay from the levels $e$
      to the levels $n,J$.
      Other off-resonant states are included as shown.
    } \label{fig:Lambdasystem}
  \end{center}
\end{figure}

To transfer the population from the state $\vert J,n\rangle$ to $\vert
J-1,n+1 \rangle$, various coherent processes can be employed.  Here we focus
on stimulated Raman adiabatic passage (STIRAP)
\cite{Bergmann1995,Bergmann1998,Vitanov2001}, 
Stark chirped Raman adiabatic passage
(SCRAP or SIARP)
\cite{Grischkowsky1975,Yatsenko99,Rickes2000},
and chirped adiabatic rapid passage (CARP)
\cite{Melinger94,Chelkowski95,Chelkowski97,Malinovsky2001} 
which offer highly efficient population transfer in
combination with robustness against variations of the pulse parameters. In
order to simplify the numerical simulations we first investigate these
processes in a system with just two rotational levels.  
That is, we examine in detail a single step in our multi-pulse coherent
transfer scheme. In this case the most
important states are the two lowest motional states for $J=0$ and the lowest
($n=0$) motional state for $J=1$ (see Fig.\ \ref{fig:Lambdasystem}). 
However,
to include the off-resonant effects all nine of the states shown in Fig.
\ref{fig:Lambdasystem} are included in the numerical calculation.  We choose
parameter ranges which are relevant for an experimental implementation. The
electronically excited state can decay to both rotational states as
well as into the uncoupled state $\vert u\rangle$.

\subsection{Stimulated Raman adiabatic passage} \label{sec:31}

STIRAP is widely used in the optical control of molecules
\cite{Vitanov2001,Rice,Kral07} where a Stokes pulse and a pump pulse are used in a
``counter-intuitive'' order to transfer the population between two
states.  The main requirement for STIRAP is the two-photon resonance
condition which corresponds to $\Delta^p_1(t)= \Delta^s_1(t)$ for a
transition on a motional sideband (see Fig.\ \ref{fig:Levelsystem}).  In
addition to this the pulse area must be large, $\Omega_0 T\gg1$, and the delay
between the Stokes and pump pulse must be of the order of the pulse width,
$\tau\approx T$, to ensure the adiabatic evolution of the system.  Another
constraint arises from the necessity to address particular motional
sidebands.  The narrow splitting of the motional states imposes the use of
laser pulses with narrow bandwidth $\nu T\gg1 $ \cite{Coulston1992}. Fast
pulses will inevitably result in the coupling of the target state to other,
close-lying states.  This in turn will reduce the transfer efficiency.
Furthermore, the resolved sideband condition \cite{Leibfried2003} requires
that the Rabi frequencies are small compared to the trap frequency, i.e.\
$\Omega_0\ll\nu$. Thus, only slow and weak pulses can be used which, as we
will see, substantially reduces the efficiency of STIRAP.  However, this can
be overcome by employing other adiabatic passage schemes as described below.

\begin{figure}
  \begin{center}
   \subfigure[]{
    \includegraphics[width=\myfigwidth]{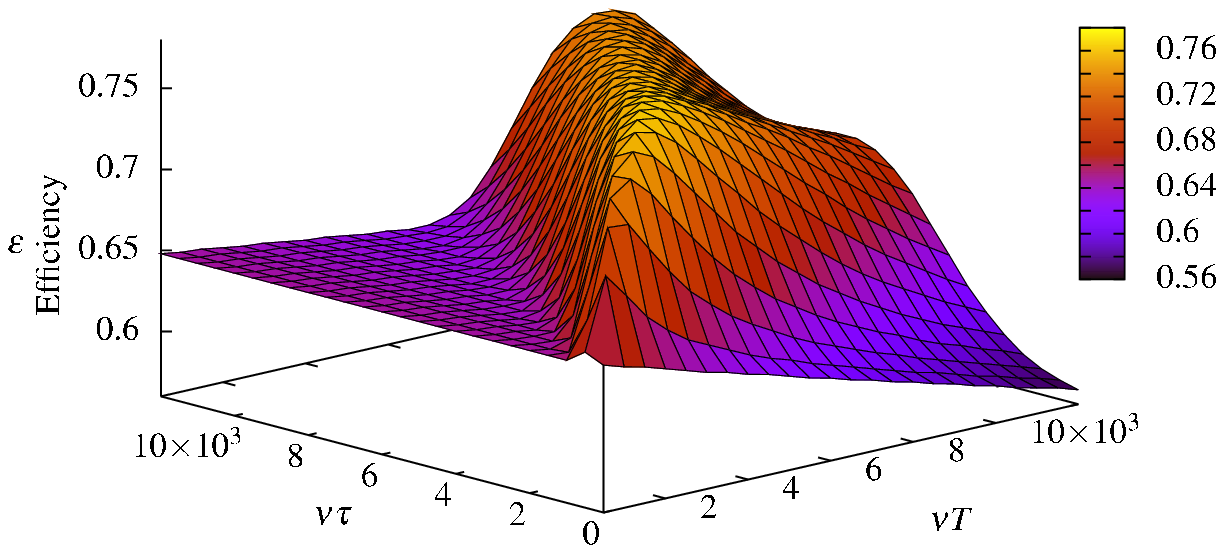}
    \label{fig:STIRAPa}}
    \\ 
    \subfigure[]{
      \includegraphics[width=\myfigwidth]{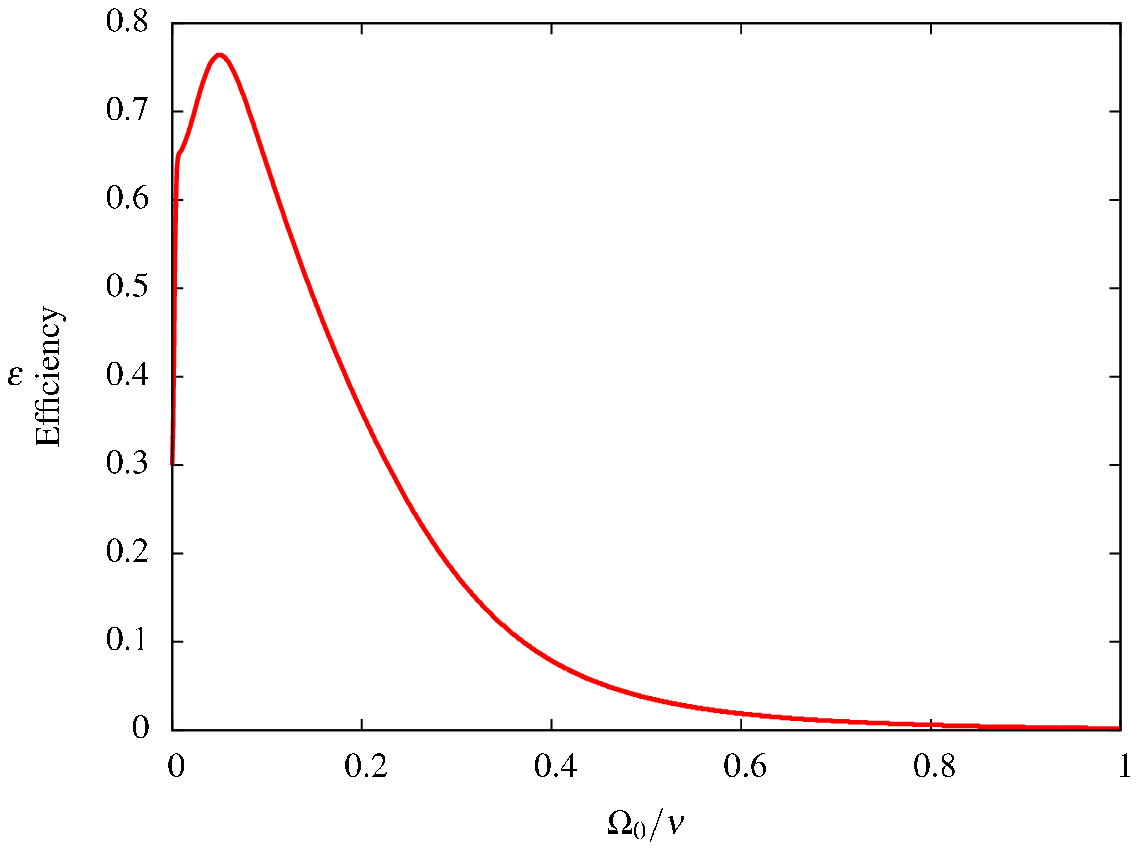}
      \label{fig:STIRAPb}}
    \caption{  \label{fig:STIRAP}    (Color online) 
      (a) The population transfer for STIRAP with different delay
      times $\tau$ and pulse widths $T$. The calculations are for the $\Lambda$-system in
      Fig.\  \ref{fig:Lambdasystem}. The Rabi frequency is
      $\Omega_0=\nu/20$. 
      (b) The efficiency as a function of Rabi frequency for
        $T=4500/\nu$, $\tau=3500/\nu$.  
      In (a) and (b) the
      pump and Stokes pulse detunings are $\Delta^p_{\mysize1}=
      \Delta^s_{\mysize1}=0$ and the other parameters are: 
      $\Gamma_J=\Gamma_u=0.01\nu$,
      $\eta=0.1$, $P_{\mysize 0,0}(-\infty)=0.3$ and 
      $P_{\mysize 1,0}(-\infty)=0.7$.
    }
  \end{center}
\end{figure}

In Fig.\ \ref{fig:STIRAPa} the transfer efficiency $\epsilon$ is plotted for
different pulse widths $T$ and delay times $\tau$.  As mentioned above the
initial mixed state is described by the populations $P_{0,0}(-\infty)=0.3$
and $P_{1,0}(-\infty)=0.7$.  The chosen decay rate of $\Gamma = 0.01 \nu$ is
in the typical range of values for the decay rate of an electronically
excited molecule, when compared to a typical trap frequency $\nu$ (of
the order of several MHz).  The results in Fig.\ \ref{fig:STIRAPa} show that
the efficiency $\epsilon$ is below 76\%. The performance of STIRAP for short
pulses is relatively poor due to the violation of the adiabaticity requirement
($\Omega_0 T \gg 1$) and the limitation on the Rabi frequency $\Omega_0$
posed by the resolved sideband condition ($\Omega_0 \ll \nu$).  For long
pulses the efficiency of STIRAP is compromised by the 
spontaneous decay $\Gamma_{u,J}$, so
the compensation of small Rabi frequencies by long laser pulses is not an
option. Therefore, the efficiency of STIRAP in this parameter range is 
low and the population transfer is governed by optical pumping rather than
coherent evolution. This is particularly visible in the plateau region for
$\tau \gg T$. In this regime the pulse delay is too large to sustain the
adiabatic evolution of the system, which results in a net loss of the ground
state population.  Highly efficient, fast STIRAP between motional states
requires large motional frequencies which are beyond current ion-trap
technology.  

\begin{figure}
  \begin{center}
    \includegraphics[width=\myfigwidth]{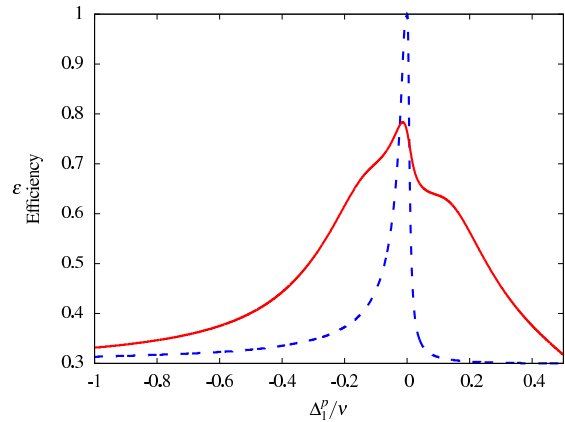}
    \caption{(Color online) Efficiency of the STIRAP process as a function of detuning
      $\Delta^p_{\mysize1}$ ($=\Delta^s_{\mysize1}$).
      We show both the case with decay,
      $\Gamma_J=\Gamma_u=0.01\nu$ (solid line), 
      and without decay,
      $\Gamma_J=\Gamma_u=0$ (dashed line).
      The other parameters are as in 
       Fig.\ \ref{fig:STIRAP} with the best values taken for 
      $T=4500/\nu$, $\tau=3500/\nu$, and $\Omega_0=\nu/20$.      
    } \label{fig:STIRAP-DETUNING}
  \end{center}
\end{figure}

We can try to suppress the excited-state population by detuning the Raman
transition from the excited state.  However, it has been known for some
time that detuning adversely affects the STIRAP process in the absence of
decay \cite{vitanov97a}. With decay present one has to consider the balance
of the adverse effect of detuning against a possible reduction in
spontaneous emission from the excited state of a model $\Lambda$ system.
Studies with such systems support the suggestion that the minimal losses
(for moderate decay rates) are found by remaining on resonance
\cite{vitanov97b,Romanenko97,Vitanov2001}.  Figure
\ref{fig:STIRAP-DETUNING} shows how the efficiency of STIRAP drops, for our
model system, as we detune from resonance.  The parameters are those for
Fig.\ \ref{fig:STIRAPb}, with the Rabi frequency $\Omega_0$ chosen to be at
the peak of efficiency in Fig.\ \ref{fig:STIRAPb}. We see that both with,
and without, decay processes it is best to be resonant. In the case
$\Gamma_J=0$ the resonance is much sharper, however.

To exploit the adiabatic evolution of the system and to suppress the
excited-state population by far detuning the Raman transition leads us to
Stark chirped Raman adiabatic passage (SCRAP) and chirped adiabatic rapid
passage (CARP), which will be discussed in Sections \ref{sec:32} and
\ref{sec:33}.  In the limit of far detuning, the excited state population
is strongly suppressed and the system's dynamics are effectively that of a
two-level system \cite{ionraman}.  The effective Raman coupling between the
states $\vert J,n\rangle$ and $\vert J-1,n+1\rangle$ is
\begin{equation} \label{eq:12}
  \Omega_{\mysize J}(t)=\frac{\eta\Omega^p_{\mysize J}(t)\Omega^s_{\mysize
      J}(t)}{\Delta^p_{\mysize J}(t)},
\end{equation}
and the effective splitting of the coupled rotational levels is 
\begin{equation} \label{eq:13}
  \Delta_{\mysize J}(t)=\delta_{\mysize J}(t)+S^s_{\mysize J}(t)-S^p_{\mysize J}(t)
   ,
\end{equation} 
with the effective two-photon Raman detuning $\delta_{\mysize J}(t)=\Delta^s_{\mysize J}(t)-\Delta^p_{\mysize J}(t)$, 
and the two Stark shifts 
$S^s_{\mysize J}(t)=[\eta\Omega^s_{\mysize J}(t)]^2/4\Delta^s_{\mysize
J}(t)$ and $S^p_{\mysize J}(t)=[\Omega^p_{\mysize J}(t)]^2/4\Delta^p_{\mysize
J}(t)$ induced by the Stokes and pump pulse respectively.

Within this effective two-level model, adiabatic rapid passage techniques
(ARP) \cite{Malinovsky2001} can be applied.  The main idea behind ARP is to
drive the system through the resonance ($\Delta_J =0$) adiabatically, to
achieve a complete population transfer.  The technique of Stark chirped
Raman adiabatic passage (SCRAP)
\cite{Grischkowsky1975,Yatsenko99,Rickes2000} takes advantage of the Stark
shifts whereas the chirped adiabatic rapid passage (CARP)
\cite{Melinger94,Chelkowski95,Chelkowski97,Malinovsky2001} uses overlapping
laser pulses along with frequency chirps to transfer the population. We
turn to these methods in the next sections.

\subsection{Stark chirped Raman adiabatic passage} \label{sec:32}

For the case of SCRAP (earlier known as self-induced adiabatic passage, or
SIARP \cite{Grischkowsky1975}) the laser pulses are engineered so that the
system undergoes an avoided level crossing ($\Delta_J=0$) near a maximum of
the effective coupling $\Omega_{\mysize J}(t)$ induced by the Stark shifts
due to the delay of the pump and Stokes pulses.  In order to obtain an
efficient population transfer the system needs to evolve adiabatically in
the crossing region \cite{Rickes2000}. For Gaussian pulses this leads to
the condition:
\begin{equation}\label{eq:14}
\Omega_0^2 T^2 \gg \left|\tau\right|\left|\Delta_J^p\right|\exp\left(2\tau^2/T^2\right).
\end{equation}

\begin{figure}
  \begin{center}
    \subfigure[]
              {\includegraphics[width=\myfigwidth]{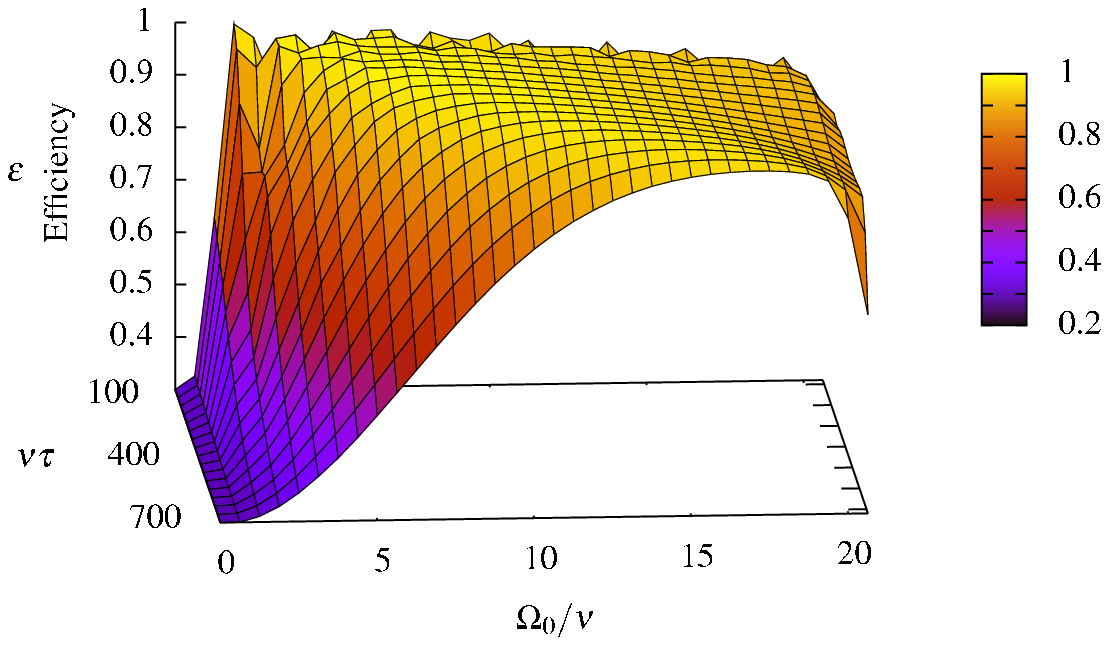}\label{fig:SIARP_a}}
              \\
              \subfigure[]{
                \includegraphics[width=\myfigwidth]{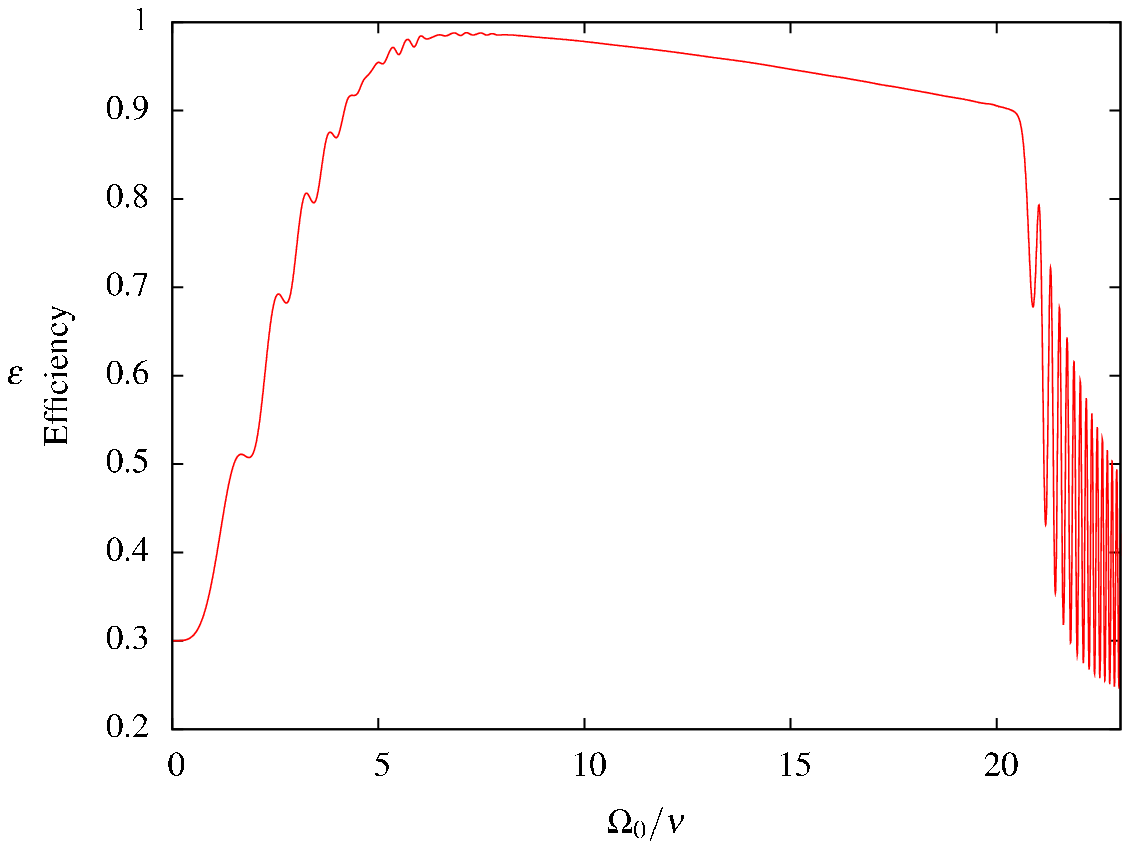}\label{fig:SIARP_b}}
              \caption{(Color online) (a) The efficiency for SCRAP for different Rabi frequencies
                $\Omega_{\mysize0}$ and delay times $\tau$ for the $\Lambda$-system in  
                Fig.\ \ref{fig:Lambdasystem}. The pulse width is $T=800/\nu$ and the pump 
                and Stokes pulse detunings are
                $\Delta^p_{\mysize1}=
                \Delta^s_{\mysize1}= 100\nu$.
                (b) The efficiency for a delay $\tau=320/\nu$. Other 
                parameters are as in Fig.\ \ref{fig:STIRAPa}.} \label{fig:SIARP}
  \end{center}
\end{figure}
Figure \ref{fig:SIARP_a} shows the efficiency $\epsilon$ for various Rabi
frequencies and pulse delays for a decay rate of $\Gamma = 0.01 \nu$ and a
fixed pulse length of $T=800/\nu$. The behavior described by the
adiabaticity requirement is clearly visible. For a fixed Rabi frequency the
efficiency decreases with increasing pulse delay $\tau$ as predicted by
Eq.\ (\ref{eq:14}). The improvement due to increased Rabi frequency is also
evident. However, for large Rabi frequencies the pulse violates the
resolved sideband condition $\Omega_{\mysize J}(t) \ll \nu$, leading to a
sudden deterioration of the efficiency $\epsilon$ at large intensities, see
Fig.\ \ref{fig:SIARP_b}. For small Rabi frequencies the efficiency
$\epsilon$ strongly depends on the delay times. For $\tau<80/\nu$ the
method is not robust. As the effective decay rate increases with increasing
Rabi frequencies, the efficiency slowly degrades for greater laser
intensity. This leads to a ridge in the $\tau$-$\Omega$ diagram. Because
the efficiency depends on the pulse delay and the Rabi frequency, accurate
knowledge of these pulse parameters is required. This can be moderated by
increasing the detuning of the laser pulses.  Because the constraints on
the pulse length are less severe than for STIRAP the population transfer
can be faster with SCRAP. Together with the far detuning this leads to a
improved robustness against the detrimental effect of spontaneous decay
\cite{note3}.

\subsection{Chirped adiabatic rapid passage} \label{sec:33}

Another way of achieving an adiabatic population transfer is the
application of simultaneous pump and Stokes pulses ($\tau =0 $) with one
laser having a frequency chirp. 
This is a Raman chirped adiabatic passage, 
 sometimes called RCAP \cite{Chelkowski95},
  though here we refer to it as chirped adiabatic rapid passage (in a
  $\Lambda$-system) or CARP. With this system
the Stark shifts are eliminated, and the detuning
$\Delta_{\mysize J}(t)$ reads
\begin{equation} %\label{}
\Delta_{\mysize J}(t)=\delta_{\mysize
  J}(t)= \alpha(t-(J_{max}-J)\tilde{\tau}).
\end{equation}
In order to ensure the adiabatic evolution the chirp rate $\alpha$ needs to
fulfill 
$\left|\alpha\right|\ll \Omega_J^2$ and $\left|\alpha\right|T^2\gg 1$ 
[with the two-photon Rabi frequency $\Omega_J$ given by Eq.\ \eqref{eq:12}].
These conditions arise from a Landau-Zener adiabaticity and from requiring
completion of a Landau-Zener transfer within the time-scale of the pulse.
Because there is no limit on the pulse duration arising directly from
the adiabaticity requirements the transition can be fast: it is only limited
by the narrow bandwidth condition $\nu T \gg 1$ \cite{Malinovsky2001}.  This
in turn reduces the susceptibility to spontaneous emission.  The resolved
sideband condition \cite{Leibfried2003} requires that $\nu\gg\Omega_{\mysize
  J}(t)$, which is easy to satisfy since the system is in the far-detuned
limit $\Delta^p_{\mysize J}\gg\Omega_{\mysize0}$.  Under these conditions
the population transferred to the target state can be estimated with the
Landau-Zener formula \cite{Landau1932a,Zener1932}
\begin{equation} \label{eq:15}
  P{\mysize J-1;n+1}(\infty)=P_{\mysize
    J,n}(-\infty)(1-e^{-\pi \Lambda_J}),
\end{equation}
\begin{figure}
  \begin{center}
    \subfigure[]
    {\includegraphics[width=\myfigwidth]{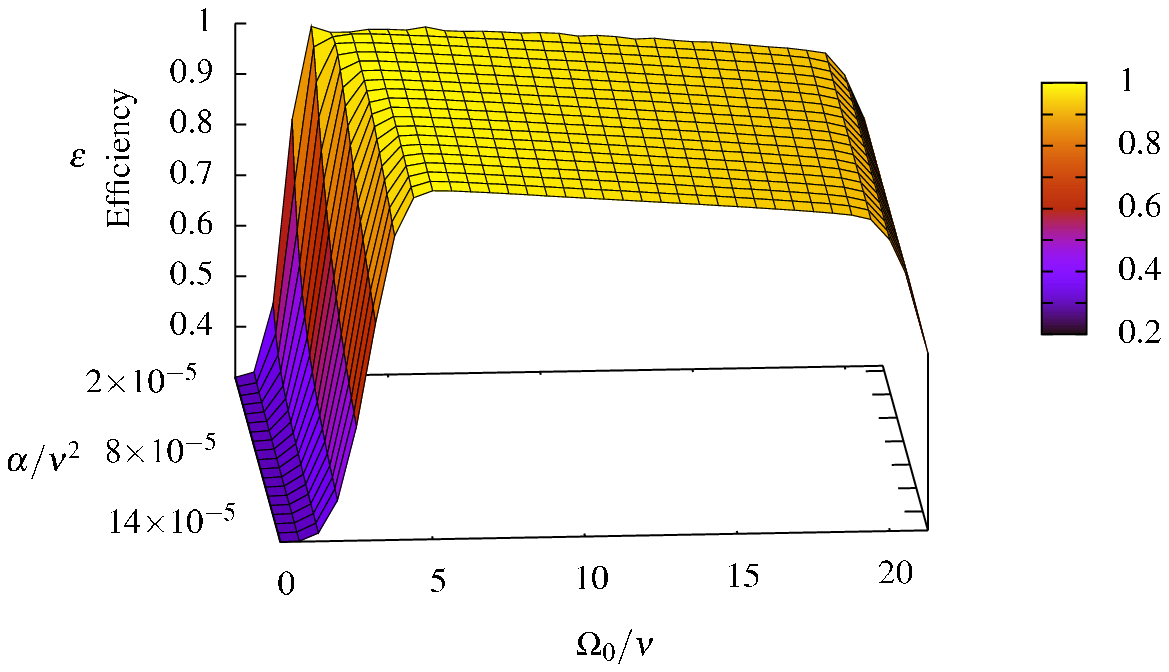}\label{fig:CARP_a}}
    \\ 
    \subfigure[]
              {\includegraphics[width=\myfigwidth]{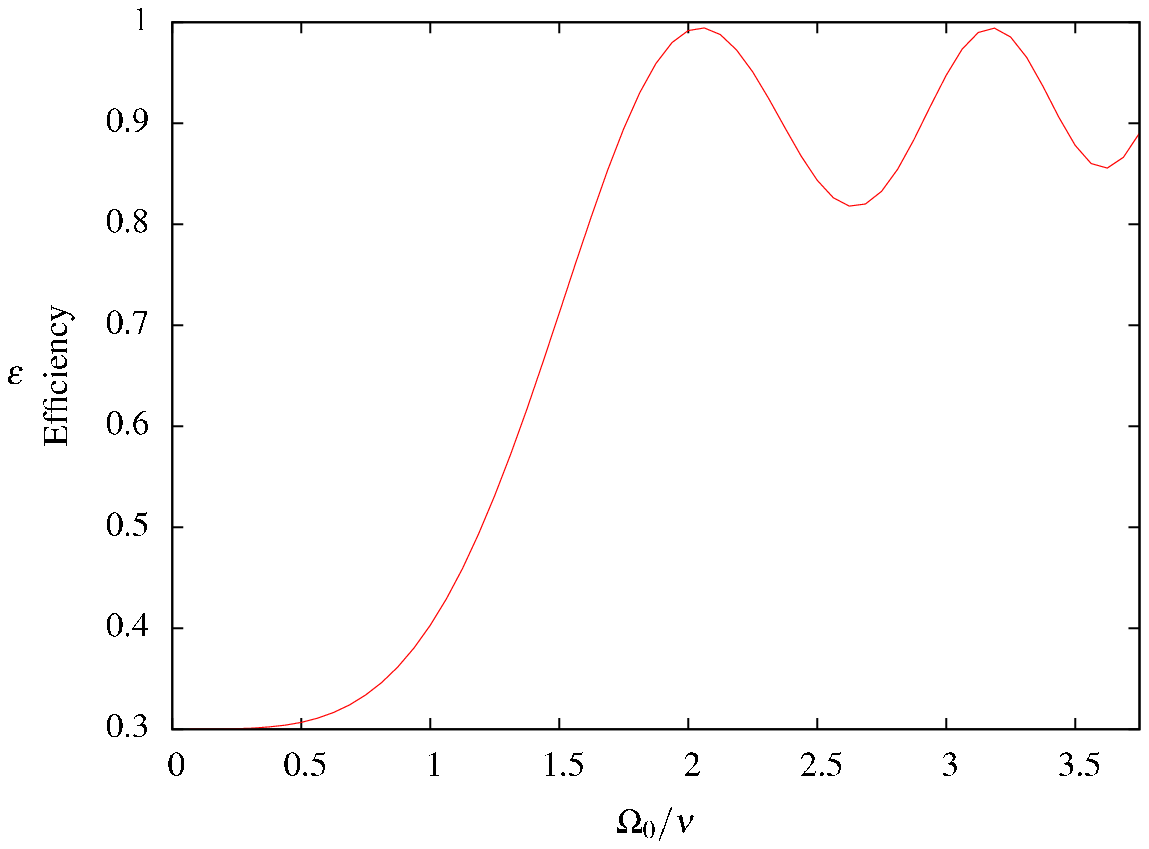}\label{fig:CARP_b}}
              \caption{(Color online) (a) The efficiency of CARP for different Rabi
                frequencies $\Omega_{\mysize0}$ and chirp rates $\alpha$ for
                the $\Lambda$-system in Fig.\ \ref{fig:Lambdasystem}. The
                delay is $\tau=0$, the pulse width is $T=800/\nu$, 
                and the Stokes pulse detuning is given by
                $\Delta^s_{\mysize1}=\Delta^p_{\mysize1} -\alpha t$. (b) The
                efficiency as a function of Rabi frequency
                  $\Omega_0$ for a small chirp rate equal to
                $\alpha=8\times10^{-6}\nu^2$. Other parameters are as in
                Fig.\ \ref{fig:STIRAPa}
                with pump pulse detuning
                $\Delta^p_{\mysize1}=100\nu$.} \label{fig:CARP}
  \end{center}
\end{figure}
where $\Lambda_J=\Omega_{\mysize J}(0)^2/2\vert\alpha\vert=\eta^2\Omega^4_{\mysize0}/(2(\Delta^p_{\mysize
  J})^2\vert\alpha\vert)$.
This behavior is confirmed by our numerical simulation [see Fig.\
\ref{fig:CARP_a}]. 
It shows the simulated efficiency for different Rabi
frequencies and chirp rates for a system initially in a state with $
P_{0,0}(-\infty)=0.3$ and $P_{1,0}(-\infty)=0.7$.
%
% fig6
%
The efficiency increases rapidly with increasing Rabi frequency $\Omega_0$
until it reaches a plateau. In this region the transfer efficiency is well
above 98\%. For small Rabi frequencies the efficiency $\epsilon$ deteriorates with
increasing chirp rate. However, this effect diminishes for large pump
intensities. Similarly to SCRAP, the increase in the effective decay rate for
high Rabi frequencies leads to a slow degradation of the transfer efficiency
for large pulse intensities. For very high laser intensities the efficiency
rapidly drops due to the violation of the resolved sideband condition
$\Omega_J(0)\ll\nu$. In a different parameter regime, that of small chirp rates
$\alpha\ll10^{-5}\nu^2$, the efficiency oscillates with changing Rabi
frequency, see Fig.\ \ref{fig:CARP_b}. Here the evolution is governed by Rabi
oscillations between the two rotational states which together with the
finite pulse length leads to large fluctuations in the efficiency. In this
regime a very precise control of the laser pulses is required, so CARP is not
robust for very small chirp rates.  For large Rabi frequencies, the
efficiency is essentially independent of the chirp rate and
CARP provides the best robustness against uncertainties in the
pulse parameters.  

\subsection{Comparison of the adiabatic passage methods}

\begin{figure}
  \begin{center}
    \includegraphics[width=\myfigwidth]{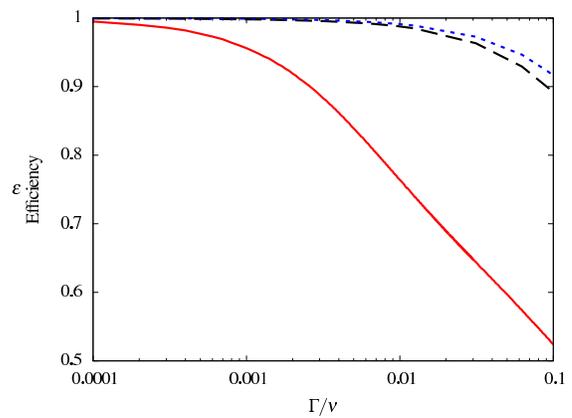}
    \caption{(Color online) The population transfer for the
      $\Lambda$-configuration of Fig.\
      \ref{fig:Lambdasystem}, for different decay rates
      $\Gamma=\Gamma_J=\Gamma_u$
      and the three methods we consider. The
        parameters are optimised at 
      $\Gamma_J=\Gamma_u=0.01\nu$ for
        each of the three
        methods and are listed separately in the
        following.
      For STIRAP (solid line):
      $\Omega_{\mysize0}=\nu/20$, $T=4500/\nu$, $\tau=3500/\nu$, and
      $\Delta^p_{\mysize1}=\Delta^s_{\mysize1}=0$.
      For SCRAP (dashed line): $\Omega_{\mysize0}=7.5\nu$,
      $T=800/\nu$, $\tau=320/\nu$, and
      $\Delta^p_{\mysize1}=\Delta^s_{\mysize1}= 100\nu$.
      For CARP (dotted line): $\Omega_{\mysize0}=5\nu$, $T=800/\nu$,
      $\tau=0$, $\Delta^p_{\mysize1}=100\nu$,
      $\Delta^s_{\mysize1}=\Delta^p_{\mysize1} -\alpha t$ with $
      \alpha = 4.69 \times 10^{-5}\nu^2$.  
      Other parameters which are fixed for all three cases are:
      $J_{max}=1$, $n_{max}=2$, $\eta=0.1$, 
      $P_{0,0}(-\infty)=0.3$, and
      $P_{1,0}(-\infty)=0.7$.  } \label{fig:Comparison}
  \end{center}
\end{figure}

Both of the adiabatic rapid passage methods, SCRAP and CARP, provide fast
population transfer. Together with the large detuning of the Raman
transition from the excited state they provide robustness against the
adverse effect of spontaneous emission.  This is clearly evident in Fig.\
\ref{fig:Comparison} where the efficiency of the three methods is plotted
against the decay rate.  
In this figure, the values of the parameters
$\Omega_0$, $T$ and $\tau$ are optimised for $\Gamma=0.01\nu$ and the
detunings given.  (In the case of CARP, the value of $\alpha$ is also
optimal.)  The same optimal values have been used for fixed parameters in
Figs.\ \ref{fig:STIRAP}--\ref{fig:CARP}.  The parameter $\Gamma$ has been
fixed to a reasonable value for diatomic molecules like N$_2^+$ or CO$^+$.
This optimisation gives a fairly wide range of parameters in Fig.\
\ref{fig:Comparison}.

Fig.\ \ref{fig:Comparison} also shows that, even though STIRAP is efficient
for small decay rates, it decreases rapidly for larger decay rates.  SCRAP
and CARP are far more efficient and their suppression of the excited-state
population exceeds that of STIRAP within the limits imposed by the
experimental requirements.  Even though SCRAP and CARP exhibit similar
efficiencies CARP is more robust against uncertainties in the laser pulse
parameters. Hence, we use CARP for our investigation of the population
transfer in a multilevel system with $J_{max}=5$ in the next Section.

\section{Chirped adiabatic rapid passage in a multilevel system ($J_{max} > 1$)} \label{sec:4}

\begin{figure}[!]
  \begin{center}
    \subfigure[~Initial population distribution.]
              {\includegraphics[width=\myfigwidth]{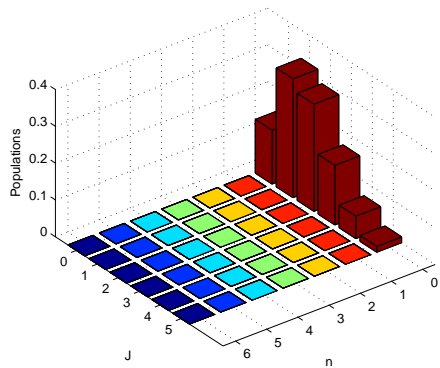}\label{fig:Multilevel1}}
              \\
              \subfigure[~Final population distribution.]{\includegraphics[width=\myfigwidth]
                {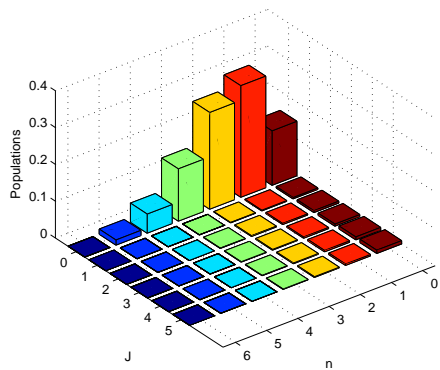}\label{fig:Multilevel2}}
              \caption{(Color online) The initial thermal distribution (a) for a system with $J_{max}=5$ and
                $\beta B=0.15$, and the final population distribution (b) after the
                completion of five sequential chirped two-photon Raman
                transitions. The detuning for the pump pulse is equal:
                $\Delta^p_{\mysize J}=100\nu$, the Rabi frequency is
                $\Omega_{\mysize0}T=5\nu$, and all the decay rates are equal to
                $\Gamma_{\mysize J}=\Gamma_{\mysize u}=0.01\nu$. Other parameters: $\eta=0.1$, $\tau=0$,
                $\tilde{\tau}=4800/\nu$, $\alpha=16\times10^{-5}\nu^2$, $T=800/\nu$ and $n_{max}=6$.}
              \label{fig:Multilevel}
  \end{center}
\end{figure}

Having derived the conditions for efficient population transfer with
chirped two-photon Raman transitions, we turn now to systems where we
include a larger number of rotational states in the calculation
($J_{max}>1$).  Consequently the mapping process involves multiple
($J_{max}$) pairs of laser pulses.  We investigate the state mapping given
in Eq.(\ref{eq:6}), for a thermal initial state distribution with $\beta
B=0.15$. This corresponds to a system where approximately six rotational
levels are significantly populated and consequently we will take
$J_{max}=5$.  In our simulation we use a Lamb-Dicke parameter of $\eta=0.1$
and a decay rate of $\Gamma=0.01\nu$ which for realistic trap frequencies
(of the order of a few MHz) corresponds to typical decay rates of
electronically excited states of diatomic molecules.  The pump laser is
detuned by $\Delta^p_{\mysize J}=100\nu$, and the Stokes laser is chirped
with the rate $\alpha=16\cdot 10^{-5} \nu^2$. Both lasers have a peak Rabi
frequency of $\Omega_{\mysize0}= 5 \nu$, a pulse length of $T=800/\nu$ and
a delay between successive pulse pairs of $\tilde{\tau}=4800/\nu$.  These
parameters were chosen on the basis of the simulations with two rotational
levels (Section \ref{sec:3}) and realistic experimental parameters.

\begin{figure}
  \begin{center}
    \includegraphics[width=\myfigwidth]{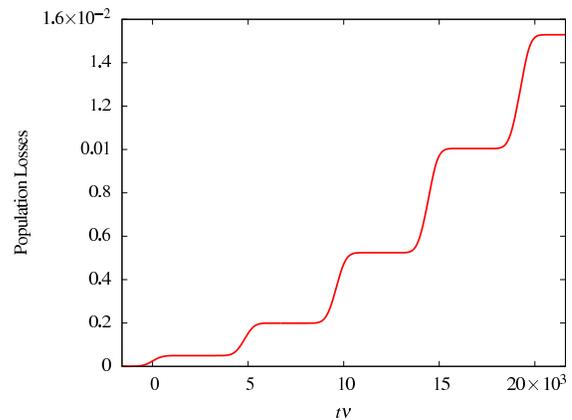}
    \caption{(Color online) The (small) population loss as a function of time for the
      two-photon chirped Raman passage in a system with $J_{max}=5$,
      see Fig.\ \ref{fig:Multilevel}.  The detuning for the pump pulse
      is equal to $\Delta^p_{\mysize1}=100\nu$ and the Rabi frequency is
      $\Omega_{\mysize0} = 5 \nu$.  
      Other parameters are as in Fig.\ 
      \ref{fig:Multilevel}.} \label{fig:Losses}
  \end{center}
\end{figure}

In Fig.\ \ref{fig:Multilevel1} the initial population distribution over all
states $\vert J,n\rangle$ is plotted, along with the final distribution in
Fig.\ \ref{fig:Multilevel2}. Apart from some small population loss, see Fig.\ 
\ref{fig:Losses}, and some weak scattering of population into states other
than the states $\vert 0,n\rangle$, the two distributions agree very well.
The total population transferred into the rotational ground-state is $92\%$,
while the total population loss into the uncoupled states is only $1.5\%$.

The remaining 6.5\%\ of the total population is mostly left in the initial
states $\vert J>0,0\rangle$ [see Fig.\ \ref{fig:Multilevel2}].  Losses due
to spontaneous emission can be further reduced by increasing the detuning.
The population which remained in the higher, coupled rotational states
$\vert J>0,0 \rangle$ can be transferred into the ground-state by
sympathetically cooling the molecule's motion and reapplying the cooling
pulse sequence. Using this approach the population of the ground-state can
be even further increased from 92\% to above 98.4\% with a total loss into
the uncoupled state of only 1.6\%.  Additional simulations with
$\Delta^p_{\mysize J}=200\nu$ and $10^3\nu$, resulted in higher
efficiencies, $95.7\%$ and $98.9\%$, respectively. The corresponding losses
are less than $0.9\%$ for $\Delta^p_{\mysize J}=200\nu$ and $0.2\%$ for
$\Delta^p_{\mysize J}=10^3\nu$.

For large numbers of populated states the transfer efficiency can be
estimated with the following equation:
\begin{equation} \label{eq:16} P_{total}
  \approx\sum_{J=0}^{J_{max}}P(J_{max}-J)\epsilon^{J_{max}-J},
\end{equation}
where $\epsilon$ is the transfer efficiency for the simple $\Lambda$-system,
and $P(J)$ is the initial population distribution (\ref{eq:5}). This agrees
well with our numerical simulation of six rotational levels.

Starting from a vibrationally cold system, these values of the efficiency
$\epsilon$ show that we can reach the motional and rotational ground-state
of molecules \cite{note2}.  For a CO$^+$ ion with a rotational temperature
of $T\approx100$K, the first 15 rotational levels are significantly
occupied initially with a ground-state population of only 3\%. By applying
the CARP state mapping with a detuning of $\Delta^p_{\mysize J}=100\nu$ and
taking the selection rule $\Delta J = 0,\pm 2$ into account the population
of the two lowest lying states can be can be increased to 85\%. Using a
detuning of $\Delta^p_{\mysize J}=10^3\nu$ this can be improved to 97.8\%
for a single cooling cycle. For a trap frequency of 4~MHz this cooling cycle
will be completed within 10~ms.

\section{Conclusion} \label{sec:conclusion}

In this work we presented an efficient method to cool the internal states
of molecules by means of coherent processes (and sympathetic cooling) thus
suppressing the problematic spontaneous decay into uncoupled states.  By
coupling the internal molecular state to the motion of the molecule, that
internal state can be mapped onto a motional state. Utilizing this, the
\emph{internal} state is cooled close to its ground-state if the molecule's
motion was initially reduced to the motional ground-state through
sympathetic cooling.  Ultimately all the degrees of freedom of the molecule
can be cooled by the application of sympathetic cooling to the final
motional excitation.  Due to its high efficiency the method presented here
is not only useful to cool the internal state, but can also be employed to
detect the internal state of the molecule by measuring its motional state
with an atom which is trapped alongside.

We have studied various adiabatic methods for a range of laser
pulse parameters which are relevant for an experimental
implementation of this cooling scheme.  The motion of the ion
imposes restrictions on the dynamics of the population transfer
process which severely limit the possible parameter range for the
laser pulses.  For the near-resonant method (STIRAP), population
transfer efficiency is very low accompanied with a large
population of the excited state.  Population losses can be
suppressed, if far-detuned chirped adiabatic two-photon Raman
passage methods are employed. Schemes that use chirped laser
pulses (CARP), or self-induced adiabatic passage (SCRAP/SIARP) by
Stark shifting the transition frequencies, turned out to be very
efficient. When it comes to the comparison of CARP and SCRAP the
former method has the advantage of easy optimization since it has
no dependence with respect to pulse shape.  Furthermore, for both
methods, and unlike STIRAP, the resolved sideband condition
imposes less severe constraints on the useful parameter space.

The requirements for all three methods were derived with simulations for a
$\Lambda$-system. Using the results from this simple model, we were able to
demonstrate the applicability of CARP in systems with more than two
rotational states. For far-detuned transitions, a high-fidelity population
mapping from the internal to the motional degrees of freedom is possible.
Losses were very low and our simulations indicate that the fidelity can be
further improved by detuning the laser pulses further from the transition.

In the scheme we propose here, each rotational level is coupled to the
excited state by a laser. Due to the large rotational level splitting of
light molecules this means in turn that multiple lasers are required. The
number of levels $N$ which have a population larger than the cut-off
population $P$, and therefore the number of required lasers, can be
estimated as $ N\approx\sqrt{-\ln(2P) k_B T / B}$ for small $P$. Even
though the number of levels $N$ for molecules at room temperature can be of
the order of 25, for temperatures of a few Kelvin this reduces to well
below 10 states. In many experiments molecular ions can be prepared in low
lying rotational states by employing photo-association or state selective
photo-ionisation in conjunction with supersonic beam expansion or
buffer-gas cooling. However, due to the interaction with black-body
radiation and collisions the internal temperature quickly thermalises. By
applying the scheme proposed here, this thermalisation can be suppressed to
maintain the ground-state population.  Additionally, the state mapping can
be employed to detect the internal states of the molecule in a
non-destructive manner which is beneficial for high-resolution spectroscopy
of molecules.

In conclusion, we have developed a fast scheme for cooling the internal states
of single molecules by employing adiabatic passage methods which provide a
high efficiency in conjunction with robustness against variations in the
parameters of the involved laser pulses.
 
\acknowledgments

We would like to thank Prof.\ W.\ Lange for many helpful comments.
This work has been supported in part by the European
  Commission's ITN project FASTQUAST.


\begin{thebibliography}{99}
\bibitem{note3} As explained in Ref.\
  \cite{Shore92}, the far off-resonant
  STIRAP process can be seen as a limit of SCRAP/SIARP
  in the absence of decay. In the presence of decay the
  STIRAP process is affected more due to the involvement
  of the excited state in the dynamics taking place
  over a long time.
\bibitem{note2} 
Due to the generally different rotational constants, any
    population in vibrationally excited states is basically
    unaltered by the mapping process we have presented here. Thus only the
    vibrational ground-state of the molecule is transferred into its
    rotational ground-state.
\end{thebibliography}
\end{document}